\documentstyle[12pt]{article}
\begin{document}

\vskip 3.0cm

\renewcommand{\thefootnote}{\fnsymbol{footnote}}

{\large\centerline{\bf Estimate for the $0^{++}$ glueball mass in QCD}
}

\vskip 0.8cm

\centerline{ Xiang-Qian Luo, and Qi-Zhou Chen}

\vskip 0.8cm

\centerline{\it CCAST (World Laboratory), Beijing 10080, China}

\centerline{\it Department of Physics, 
Zhongshan University, Guangzhou 510275, China 
\footnote{Mailing address}
}

\vskip 3.0cm

\centerline{\bf Abstract}
We obtain accurate result for the lightest glueball mass
of QCD in 3 dimensions from lattice Hamiltonian field theory.
Using the dimensional reduction argument, 
a good approximation for confining theories, 
we suggest that the $0^{++}$ glueball mass in 
3+1 dimensional QCD be about $1.71$ GeV.

\vfill
\eject

The spectroscopy of QCD in the pure gauge sector, 
i.e., the glueball masses attracts considerable attention.  
In the quenched approximation,
these glueball are non $q\bar{q}$ gluonic bound states 
formed by strong self-interactions of the gluons, and their masses
vary from about 1.4 Gev to 2.5 Gev.
The flux tube models, bag models, sum rules and lattice techniques
have been used to extract the glueball masses, among which  
lattice QCD seems to give more reliable estimates.
Experimentally, 
a lot of glueball candidates such as 
$\iota(1440)$, $f_0(1520)$, $\theta /f_J(1720)$ and $\xi(2230)$, 
produced in the  
$J/\psi$ radiative decays \cite{BES,Chao,Huang} are within this range. 
The difficulty in experimental identification of a
glueball comes from the complexity 
in determining the quantum numbers $J^{PC}$ of these particles. 

The Monte Carlo simulation of lattice QCD on the Euclidean lattice
has become a powerful and conventional method in spectrum calculations. 
Concerning the lightest glueball $0^{++}$,
numerical data of 15 years ago, on rather small 
lattices with poor signal to noise ratio
suggested $M(0^{++})\approx 1$ Gev. 
The value of $M(0^{++})$ seems increasing with the lattice volume. 
Most recently, more accurate calculations 
by the IBM group \cite{IBM}
on much larger lattices, higher statistics and better algorithm gave 
$M(0^{++})\approx 1.740 \pm 0.071$ Gev, where the infinite volume  
extrapolation has been made. For a review of the current status,
see ref. \cite{Wein}.

Here we would discuss an alternative way \cite{GCL,QCD3,MASS}
to extract the glueball masses and wavefunctions 
by solving the Schr{\"o}dinger equation
\cite{MASS}
\begin{eqnarray}
H \vert F \rangle = \epsilon_{F} \vert F \rangle.
\label{sch}
\end{eqnarray}
Whereas our goal is to do concrete computations \cite{Pro} in 
four dimensional QCD,
in this paper we would like to discuss quantitatively 
the properties of three dimensional $\rm{QCD}$, 
an  interesting and relevant but much simple theory.
We will then use the idea of dimensional reduction 
\cite{Green,Samuel1,Samuel2} to estimate
the $0^{++}$ glueball mass in four dimensional QCD.

With the standard notations ($g$ being the lattice gauge coupling, 
$a$ the lattice spacing, 
$U_l=exp(iga A_l)$ the gauge link, $E_l$ the color-electric field on the
link $l$, 
$U_p$ the plaquette variable), the lattice Hamiltonian is given by
\begin{eqnarray}
H= {g^{2} \over 2a} \sum_{l} E_{l}^{\alpha} E_{l}^{\alpha}
- {1
\over ag^{2}} \sum_{p} Tr(U_{p} + U_{p}^{\dagger} -2)
\label{H}
\end{eqnarray}
is the lattice version of the Yang-Mills Hamiltonian 
$\int d^{D-1} x ( e^2  \vec {E}
\cdot \vec {E} + e^{-2} \vec {B} \cdot \vec {B})/ 2$ with $e$ the
continuum gauge coupling, $\vec{E}$ and $\vec{B}$ the color electric 
and magnetic fields respectively. 
The glueball wavefunction in (\ref{sch}) is
\begin{eqnarray}
\vert F \rangle = \lbrack F(U) - 
{\langle \Omega \vert F(U) \vert \Omega \rangle
\over \langle \Omega \vert \Omega \rangle} \rbrack \vert \Omega \rangle 
\label{WAVE}
\end{eqnarray}
created by the gluonic operator $F(U)$
with the given quantum number 
$J^{PC}$ acting on the vacuum $\vert \Omega \rangle$.
The vacuum wavefunction $\vert \Omega \rangle$ satisfies 
$H \vert \Omega \rangle = \epsilon_{\Omega} \vert \Omega \rangle$.
An estimate of the glueball mass is then 
$M_{J^{PC}}=\Delta \epsilon=\epsilon_{F}-\epsilon_{\Omega}$.

In a series of papers \cite{GCL,QCD3,MASS}, 
we developed a method for solving 
the lattice Schr\"odinger equation (\ref{sch})
in a new scheme which preserves the correct continuum behavior 
at any order of approximation.
Physically, if one wants to well describe a glueball on the lattice,
the size or the Compton length of a glueball, which
is usually of the same order as that of a hadron, 
should be greater than the lattice spacing $a$.
In other words,
the low energy spectrum originates mainly 
from the long wavelength excitations.
Our starting point is to embody this physical implication 
in the eigenvalue equation when solve it approximately.
The philosophy is to have the correct long
wavelength limit at any order of approximation.
The advantage and reliability 
of such a method has been confirmed by 
the results of two-dimensional $\sigma$ models (in \cite{SIGMA}),
three-dimensional $\rm{U(1)}$ (in \cite{FLG}),
$\rm{SU(2)}$ (in \cite{GCL,CGZF,GCFC}) and $\rm{SU(3)}$ (in \cite{QCD3,MASS})
gauge theories: the results converge very rapidly, and
even at very low truncation orders clear scaling
windows for the vacuum wavefunction and mass gaps
have been established. It is very exciting that  for 
the $\sigma$ models, 2+1 D $\rm{U(1)}$,  and 2+1 D $\rm{SU(2)}$ gauge 
theories, the vacuum wavefunction and
the mass gap are in perfect
agreement with the most recent
Monte Carlo data.
In our pioneering study of 2+1 D \rm{SU(3)}, 
we obtained the first estimates for the
vacuum wavefunction \cite{QCD3} and the glueball masses \cite{MASS}.
Most of these results have been summarized in \cite{Guo,Luo}.

We begin with recapitulating briefly our method. 
Suppose the ground state has the form 
$\vert \Omega \rangle = exp \lbrack R(U) \rbrack \vert 0 \rangle$,
with $\vert 0 \rangle$ being the fluxless bare vacuum 
and $R(U)$ being a linear combination of
gauge invariant gluonic operators $G$.
Substituting this, (\ref{H}) and (\ref{WAVE}) into (\ref{sch}),
we have the following exact eigenvalue equation 
for the operator $F(U)$
\begin{eqnarray}
\sum_{l} \lbrace [E_l,[E_l,F(U)]]+ 2[E_l,F(U)][E_l,R(U)] \rbrace
={2a\Delta \epsilon \over g^2}  F(U).
\label{eig}
\end{eqnarray}
In \cite{GCL,QCD3,MASS,Guo,Luo}, 
we have illustrated how to obtain
$R(U)$ and $F(U)$ by expanding
them in
order of graphs (Wilson loops) $G_{n,i}(U)$:
\begin{eqnarray}
F(U)=\sum_{n} F_{n}(U)=\sum_{n,i} f_{n,i} G_{n,i}(U), 
\label{graph}
\end{eqnarray}
with $n$ being the order of the graphs.
In practice, equation (\ref{eig}) 
has to be truncated to some finite order $N$
\begin{eqnarray*}
\sum_{l} \lbrace [E_l,[E_l,\sum_{n}^{N} F_{n}(U)]]
+2\sum_{n_1+n_2 \le N}[E_l,F_{n_1}(U)][E_l,R_{n_2}(U)] \rbrace
\end{eqnarray*}
\begin{eqnarray}
={2a\Delta \epsilon \over g^2}  \sum_{n}^{N} F_{n}(U),
\label{c1}
\end{eqnarray}
from which the
coefficients $f_{n,i}$ are determined. 

The first term in (\ref{c1}) doesn't create higher order graphs, while
the second term generates new or higher order graphs of order $n_1+n_2$.
Therefore, one should carefully truncate the second term in this
calculation.
The essential feature of our approach 
is in the correct treatment of this second term.
It has been generally proven \cite{GCL} that in the long wavelength limit
this term should behave as
\begin{eqnarray}
  [E_l,F_{i}(U)][E_l,R_{j}(U)] 
  \propto a^6 ~Tr({\cal D} {\cal F}_{\mu,\nu})^2.
\end{eqnarray}
Not to violate this behavior, when the equation
(\ref{c1}) is truncated to the $Nth$ order, all the graphs created by 
$[E_l,F_{i}(U)][E_l,R_{j}(U)]$ for $n_1+n_2 > N$ must be discarded. 
For example,
equation (\ref{c1}) truncated to $N=2$ is
\begin{eqnarray}
\sum_{l} \lbrace [E_l,[E_l,F_1+F_2]]
+2 [E_l,F_1][E_l,R_1] \rbrace
={2a\Delta \epsilon \over g^2} (F_1+F_2).
\label{c2}
\end{eqnarray}
For N=3,
the truncated equation (\ref{c1}) is
\begin{eqnarray*}
\sum_{l} \lbrace [E_l,[E_l,F_1+F_2+F_3]]
+2 [E_l,F_1][E_l,R_1]+ 2 [E_l,F_1][E_l,R_2]
\end{eqnarray*}
\begin{eqnarray}
+ 2 [E_l,F_2][E_l,R_1] \rbrace
={2a\Delta \epsilon \over g^2} (F_1+F_2+F_3).
\label{c3}
\end{eqnarray}
Higher order truncated eigenvalue equations satisfy the same rule. 
From an eigenvalue equation at order $N$, we derive a set of 
nonlinear algebraic equations for the coefficients $f_{n,i}$
of the operators. Solving these equations, we obtain not only
the wavefunction of the glueball, but also the glueball mass
at order $N$. (For the vacuum wavefunction, see ref. \cite{QCD3}).
It is worth mentioning another advantage of such an approach: no
group integration is necessary so that higher order calculations
are feasible.
The difference between different truncation orders is the estimate 
for the systematic error in the calculation.

For a non-abelian gauge theory, 
the element $A$ of the gauge group has to satisfy the uni-modular
condition \cite{QCD3} (or Caley Hamilton relation \cite{SCHU}), 
which is for \rm{SU(3)}
\begin{eqnarray}
A_{ij} A_{kl} A_{mn}\epsilon_{jln}
=\epsilon_{ikm}.
\label{con1}
\end{eqnarray} 
Multiplying it by $A_{pi}^{\dagger}$, 
and then summing over the $i$ index,
it becomes
\begin{eqnarray}
A_{kl} A_{mn}\epsilon_{pln}
=A_{pi}^{\dagger} \epsilon_{ikm}
\label{con2}
\end{eqnarray} 
Multiplying it again by $\epsilon_{pqr}$ 
and summing it over the $p$ index, we obtain
\begin{eqnarray}
A_{il} A_{kj}
=A_{ij} A_{kl} -Tr A^{\dagger} 
(\delta_{ji} \delta_{lk} -\delta_{kj} \delta_{il})
-A_{il}^{\dagger} \delta_{kj}
+A_{ij}^{\dagger}\delta_{kl}
-A_{kj}^{\dagger}\delta_{il}
+A_{kl}^{\dagger}\delta_{ij}.
\label{con3}
\end{eqnarray} 
These formulae are useful in classification of graphics.
Because of these conditions, one should choose properly an independent 
set from the graphs generated by the
second term in (\ref{c1}). Mathematically, 
any independent set chosen in this way 
can be used in the calculation.

Physically, 
the connected graphs represent more coherence and have less
mixing with lower order graphs.
It was shown in \cite{GCFC} that the use of connected set makes
the convergence of the results much faster than the use of
disconnected set \cite{CGZF} in a
(2+1)-dimensional $\rm{SU(2)}$ model. 

For the realistic gauge group $\rm{SU(3)}$,
the complication is that not all the disconnected graphs can be
transformed
to the connected ones. 
However, we observed that
if more disconnected graphs are transformed 
according to the uni-modular conditions (\ref{con1}), (\ref{con2}) 
or (\ref{con3}) into the
connected ones \cite{QCD3}, 
the scaling behavior was much better.
We have also tested several sets of operators \cite{COM}. 
One of them is classified according to inverse of the graphs 
$G^{I}_{n,i} \propto Inv[G_{n,j}]$
generated by the second term in (\ref{c1}) with the uni-modular
conditions taken into account. (For the definition of the inverse operator, 
see ref. \cite{Green}).

Since $\rm{QCD}_3$ is a super-renormalizable gauge theory, the
renormalization requirements amounts to dimensional analysis.
In the weak coupling region (for large $\beta=6/g^2$),
because the renormalized charge $e$ and the bare coupling are related by
$g^2=e^2 a$, dimensional analysis tells us that the dimensionless masses
$aM_{J^{PC}}$
should scale as
\begin{eqnarray}
{aM_{J^{PC}} \over g^2} \to {M_{J^{PC}} \over e^2} \approx const.,
\label{scale}
\end{eqnarray}
from which the continuum
physical glueball masses $M_{J^{PC}}$ are extracted.

Using the techniques in \cite{MASS,COM} and 
after a careful analysis of our results, 
we obtain the value for $M(0^{++})/e^2$ more accurate than the 
previous paper \cite{MASS}, 
and estimate the systematic error due to the finte $N$ truncation 
to be less than 0.06.
The validity of equation (\ref{scale}) extends from $\beta=5$
to $\beta=12$, and in this range
\begin{eqnarray}
  {M(0^{++}) \over e^2} \approx 2.15 \pm 0.06,
\label{data1}
\end{eqnarray}
where the error denotes the systematic uncertainties due to the finite
order truncation [in \cite{MASS}, we obtained 
$M(0^{++})/e^2 \approx 2.1$
for $\beta \in [5,8)$ at third order approximation 
using the connected graphs as an independent basis]. 
Our value for the lightest glueball can be compared with 
Samuel's recent result \cite{Samuel2} from the 2+1 D Hamiltonian  
QCD in the continuum: $M(0^{++})/e^2 \approx 1.84 \pm 0.46$.

One may also understand the relation between the glueball mass
and the confinement scale from the vacuum wavefunction.
The vacuum functional, which interpolates the strong and weak coupling
regimes, are \cite{Arisue,Samuel2}
\begin{eqnarray}
\vert \Omega \rangle=exp \{ {1 \over 2 e^2} 
\int d^{D-1}x ~ tr [{\cal F}_{ij} 
({\cal D}_k {\cal D}_k+ \xi^{-2})^{-1/2} {\cal F}_{ij}] \},
\end{eqnarray}
with ${\cal F}$ being 
the field strength tensor in spatial dimensions 
and ${\cal D}$ the covariant derivative. 
The correlation length $\xi$, with dimension of inverse mass, 
is  proportional to $e^{-2}$, i.e.,
the confinement scale in the vacuum. It is suggested in ref. 
\cite{Samuel2} that $\xi^{-1}$ 
might also be related to the constituent gluon mass 
and the lightest glueball mass.
In the strong coupling limit or large $Nc$ (number of colors) 
limit, it reduces to the strong coupling 
wavefunction obtained by \cite{Feynman,Green},
\begin{eqnarray}
\vert \Omega \rangle=exp [ 
- {\mu_0} \int d^{D-1}x ~ tr {\cal F}^2 ].
\end{eqnarray}
In the intermediate and weak coupling, it becomes \cite{Arisue,GCL,QCD3}
\begin{eqnarray}
\vert \Omega \rangle=exp \lbrack 
- {\mu_0 \over e^2} \int d^{D-1}x ~ tr {\cal F}^2
- {\mu_2 \over e^6} \int d^{D-1}x  ~tr ({\cal D} {\cal F})^2 \rbrack,
\label{a1}
\end{eqnarray}
which is just our vacuum wavefunction 
for the long wavelength configurations \cite{GCL,QCD3}.
The correlation length has a relation with the coefficients $\mu_0$ and 
$\mu_2$:
\begin{eqnarray}
\xi=({-2 \mu_2 \over \mu_0})^{1/2}.
\end{eqnarray}
For 2+1 D SU(2), $\xi=0.65/e^2$ (see refs. \cite{Arisue,GCL}),
while for 2+1 D SU(3), our result \cite{QCD3} is $\xi=0.53/e^2$. 
If the glueball mass is proportional to the constituent gluon mass,
from the difference of the scales between SU(2) and SU(3), one 
may also guess $M(0^{++})/e^2 \approx 2$, consistent with
the result (\ref{data1}) from our practical calculation.

Combining the most recent Monte Carlo data \cite{Lut} 
for the string tension 
$\sigma$ in $\rm{QCD}_3$, which is $\surd \sigma=(0.554 \pm 0.004) e^2$, 
we obtain the ratio of the $0^{++}$ glueball mass 
over square root of the string tension in the continuum limit
\begin{eqnarray}
  {M_{0^{++}} \over \surd \sigma} \approx 3.88 \pm 0.11.
\label{data2}
\end{eqnarray}

Now we follow the argument of dimensional reduction 
\cite{Green,Samuel1,Samuel2}.
In a confining theory in $D$ space-time dimensions with $2 < D \leq 4$, 
the function in the exponential of the vacuum functional acts like 
an action of an effective field theory in $D-1$ dimensions.
In other words, in computing vacuum expectation values, 
a confining theory in $D$ dimensions 
becomes a localized field theory in $d=D-1$ dimensions. 
This can be exactly proven in the strong coupling or large $N_c$ limit.
Because in this limit, the fixed time vacuum expectation value of a
operator $O(U)$ in $D$ dimensions is
\begin{eqnarray}
{\langle \Omega \vert O(U) \vert \Omega \rangle 
\over \langle \Omega \vert \Omega \rangle} 
\to 
{\int [dU] O(U) exp[ - 2{\mu_0} \int d^{D-1}x ~ tr {\cal F}^2 ],
\over \int [dU] exp[ - 2{\mu_0} \int d^{D-1}x ~ tr {\cal F}^2 ]},
\end{eqnarray}
corresponding to the path integral expression
for $<O(U)>$ in $D-1$ dimensional lattice field theory.
It has been argued \cite{Samuel2,BF,Man} that 3+1 D theory can still 
be approximated by its 2+1 D theory
for long wavelength configurations 
in comparison to the confinement scale. According to this argument, 
$M(J^{PC})/\surd \sigma$ for the lightest glueball 
should be approximately the same
for 2+1 and 3+1 dimensions. 
Since for SU(3) the number of color is larger, and 
the measured length $\xi$ in the vacuum functional ({\ref{a1})  
is smaller than that for SU(2), our speculation is 
that the approximation is better for SU(3) gauge theory. 
In fact, equation (\ref{data2}) is consistent
with IBM data $M(0^{++})/\surd \sigma=3.95$ from
Monte Carlo simulation of 3+1 D lattice QCD, providing 
$\surd \sigma=0.44$ Gev is used. 
From this world average value 
for the string tension and (\ref{data2}),
we expect 
\begin{eqnarray}
M(0^{++})=1.71 \pm 0.05 ~ GeV, 
\end{eqnarray}
in nice agreement with the IBM
data $M(0^{++})=1.740 \pm 0.071$ \cite{IBM}.
This favors $\theta /f_J(1710)$ as a candidate of the $0^{++}$ glueball.

In conclusion, using the eigenvalue equation method developed
in \cite{GCL,QCD3}, we obtain accurate result for the lightest
glueball mass in (2+1)-dimensional QCD, with systematic uncertainty
under well control.
We also use the idea dimensional reduction to extrapolate the
results to QCD in 3+1 dimensions. Perfect agreement with
the most accurate Monte Carlo data indicates that $\rm{QCD_3}$
is not just a toy model for $\rm{QCD}_4$, and their relation
should be more deeply understood.

\vskip 2.0cm

\noindent
{\bf Acknowledgments}

We are grateful to X.Y. Fang, S.H. Guo, J.M. Liu for 
valuable collaboration in developing the eigenvalue equation method,
S. Samuel and D. Sch{\"u}tte  
for useful communications, 
X.Y. Shen and Y.C. Zhu for discussions of their 
experimental work \cite{BES} and status on searching 
$0^{++}$ glueball with the BES collaboration.
Our work was supported by the project of
the National Natural Science Foundation,
Doctoral Program Foundation, and  
National Education Committee of China.

\vfill
\eject

\vfill
\eject

\end{document}